# A Potential Cyclotron Line Signature In Low Luminosity X-Ray Sources

Robert W. Nelson[1], John C. L. Wang[2], E.E. Salpeter[3] and Ira Wasserman[3]

[1]CITA, 60 St. George St., Toronto, ON M5S 1A7, CANADA
[2]JILA, University of Colorado, Campus Box 440, Boulder, CO 80309, USA
[3]CRSR, Cornell University, Ithaca, NY 14853, USA

## ABSTRACT

Estimates indicate there may be $\gtrsim 10^3$ low luminosity X-ray pulsars ($L \lesssim 10^{34}$ erg s$^{-1}$) in the Galaxy undergoing "low-state" wind accretion in Be/X-ray binary systems, and $\sim 10^8 - 10^9$ isolated neutron stars which may be accreting directly from the interstellar medium. Despite their low effective temperatures ($kT_e \lesssim 300$ eV), low luminosity accreting neutron stars with magnetic fields $B \sim (0.7 - 7) \times 10^{12}$ G could emit a substantial fraction ($0.5 - 5\%$) of their total luminosity in a moderately broadened ($E/\Delta E \sim 2 - 4$) cyclotron *emission* line which peaks in the energy range $\sim 5 - 20$ keV. The bulk of the thermal emission from these stars will be in the extreme ultraviolet/soft X-ray regime, which is subject to strong interstellar absorption and would be difficult to distinguish from spectra of other types of objects. In sharp contrast, the *nonthermal* cyclotron component predicted here will not be strongly absorbed, and consequently it may be the only distinguishing signature for the bulk of these low luminosity sources. We propose a search for this cyclotron emission feature in long pointed observations of the newly discovered candidate isolated neutron star MS0317.7-6477, and the Be/X-ray transient pulsar 4U0115+63 in its quiescent state. We note that an emission-like feature similar to the one we predict here has been reported in the energy spectrum of the unusual X-ray pulsar 1E 2259+586.

*Subject headings:* accretion, accretion disks—line:formation— radiative transfer—stars:neutron—Xrays:stars

## 1. Introduction

Most of the 30 known X-ray pulsars in the Galaxy are bright, with luminosities $L \gtrsim L_E = 1.4 \times 10^{35} M_{1.4} R_6^{-2} A_{cap} (\sigma_T/\sigma_m)$ erg/sec, where $L_E$ is the *effective* Eddington limit for magnetic polar cap accretion onto a neutron star of mass $M = 1.4 M_{1.4} M_\odot$, radius $R = 10^6 R_6$ cm, polar cap area $A_{cap}$ km$^2$, and $\sigma_T/\sigma_m$ is the ratio of the nonmagnetic to the effective magnetic photon-electron scattering cross sections. Nearly half of these pulsars undergo wind accretion from Be star companions, but are usually detected only during high luminosity transient outbursts. These bright sources may just be the "tip of the iceberg", however, of a much larger underlying population of lower luminosity magnetic accretors. Estimates suggest there are $\gtrsim 10^3$ similar Be/neutron star binaries in the Galaxy accreting at low luminosities $L \ll L_E$ (Rappaport & van den Heuvel 1982; Meurs & van den Heuvel 1989; King 1991). In addition, assuming a production rate from supernovae of $(10 - 100 \text{ yrs})^{-1}$ there should be $\sim 10^8\text{-}10^9$ isolated neutron stars in the Galaxy, of which some fraction may be accreting directly from the interstellar medium (ISM) at detectable luminosities (Trevis & Colpi 1991; Nelson, Salpeter, & Wasserman 1991; Blaes & Madau 1993).

Although easier to detect, the emergent spectra from the high luminosity sources is extremely difficult to model owing to the strong coupling between radiation and the accretion flow (e.g. Arons, Klein, and Lea 1987). As a result, the theoretical interpretation of the observed spectra is not straightforward. By contrast, *because radiation pressure is unimportant* for low luminosity sources they could offer a much cleaner laboratory for testing the basic physics of accretion in a superstrong magnetic field environment. Here we present calculations of the emergent spectra of such low luminosity X-ray pulsars. In sharp contrast to their brighter counterparts, we find that sources with low effective temperature and field strengths $B \sim 10^{12}$ G should exhibit a unique nonthermal signature: a prominent cyclotron *emission* line, containing $0.5 - 5\%$ of the total accretion luminosity, superposed on the Wien tail of the underlying soft thermal emission. If this line can be resolved it would enable an unambiguous determination of the surface magnetic field strength, thereby providing important constraints on neutron star magnetic field evolution.

One can imagine a number of possible reasons why low luminosity accretion onto strongly magnetized neutron stars may not be able to occur, the most important of which is the centrifugal barrier at the rapidly rotating magnetosphere - the so-called propeller effect (Illarionov and Sunyaev 1975). Using standard spherically symmetric accretion theory, mass flow onto the neutron star may be inhibited when the star rotates faster than the local Kepler period at the magnetospheric boundary.

$$P_p \lesssim P_{crit} = 60 L_{34}^{-3/7} \mu_{30}^{6/7} M_{1.4}^{-2/7} R_6^{-3/7} \text{s}, \tag{1}$$

where $\mu = 10^{30} \mu_{30} \text{ G cm}^{-3}$ is the neutron star magnetic moment, and $L = 10^{34} L_{34} \text{ erg s}^{-1}$ is the luminosity from the neutron star. Equation (1) is valid for a neutron star moving relative to the local medium at speeds $v \lesssim 4 \times 10^3 \text{ km s}^{-1} M_{1.4}^{3/7} R_6^{1/7} L_{34}^{1/7} \mu_{30}^{-2/7}$, so that gravitational focusing of the accreting flow is important. It is not clear whether isolated neutron stars could have been spun down to these periods (Harding & Leventhal 1992; Blaes & Madau 1993). Nonetheless, it could be that these barriers merely reduce the accretion efficiency rather than preventing it altogether. Here we focus on what might be seen from a low luminosity, high magnetic field accreting neutron star if such systems can exist and are detectable. Clearly, the observation of a rapidly rotating ($P_p < P_{crit}$), low luminosity pulsar would necessitate some refinement of current ideas on entry of plasma into neutron star magnetospheres.

In section 2, we describe our model for magnetic accretion and radiative transport. In section 3, we present spectra based on Monte Carlo cyclotron transport simulations. In section 4, we discuss possible applications of our model to the Be/X-ray binary transients with known magnetic field strengths and to a candidate isolated magnetized neutron star. A preliminary version of these results was presented at the Maryland Conference on the Evolution of X-Ray Binaries (Nelson, Wang, Salpeter & Wasserman 1994a).

## 2. Magnetic Accretion And Radiative Transport

The physics of accreting matter decelerating in magnetized neutron star atmospheres is described in detail in Nelson, Salpeter, & Wasserman (1993) (hereafter NSW93). In strong magnetic fields, $B \sim 10^{12}$G, atmospheric electrons move freely along the field, but are quantized into Landau levels of energy $E_n = (n + 1/2)E_B$ perpendicular to the field, where

$$E_B \equiv \frac{\hbar e B}{m_e c} = 11.6 B_{12} \text{ keV}. \qquad (2)$$

Since radiative deexcitation greatly dominates collisional excitation for typical atmospheric conditions and since $k_B T \ll E_B$, atmospheric electrons normally reside in the Landau ground state ($n = 0$), and therefore constitute a sort of one dimensional gas (see, e.g., Ventura 1973). Consequently, the manner in which infalling matter decelerates and converts the bulk of its kinetic energy into radiation is dramatically different from what occurs in non-magnetic, three-dimensional atmospheres.

Accreting protons, channeled along magnetic field lines, free fall into the plasma atmosphere at the magnetic polar cap and decelerate by undergoing Coulomb collisions with atmospheric electrons. In *nonmagnetic* atmospheres, these protons would penetrate to a stopping optical depth (Thomson units) $\tau_s = m_p v_{ff}^4 / (6 m_e \ln \Lambda_c) = 5.1 M_{1.4}^2 R_6^{-2} (10/\ln \Lambda_c)$, where $v_{ff} = 0.64c\, M_{1.4}^{1/2} R_6^{-1/2}$ is the proton free fall velocity and $\ln \Lambda_c \sim 10$ is the Coulomb logarithm for a nonmagnetic atmospheric plasma. In this case, all but a fraction $\sim 1/\ln \Lambda_c$ of the initial proton energy is lost to electron recoil motion *transverse* to the proton velocity. In magnetic atmospheres, where electron motion perpendicular to the magnetic field lines is constrained, collisions with center of mass energy $m_e v^2/2 > E_B$ can result in $0 \to n$ Landau excitations up to a maximum level

$$n_{max} \equiv \frac{m_e v^2}{2 E_B} = 9.0 \left(\frac{v}{v_{ff}}\right)^2 B_{12}^{-1} M_{1.4} R_6^{-1}. \qquad (3)$$

NSW93 have shown that in moderately strong fields, where $n_{max}$ is fairly large, the stopping power of the plasma is well approximated by the nonmagnetic result, but with a reduced Coulomb logarithm,

$$\ln \Lambda_c \to \ln 2n_{max}. \tag{4}$$

In particular, accreting protons penetrate the atmosphere to a magnetic stopping depth $\tau_B \approx \tau_s(\ln \Lambda_c / \ln 2n_{max}) = 51 M_{1.4}^2 R_6^{-2} / \ln 2n_{max}$ and lose all but a fraction $\sim 1/\ln(2n_{max})$ of their initial energies in exciting atmospheric electrons to $1 \leq n \leq n_{max}$ Landau levels. The decelerating protons do not veer appreciably from their original direction of motion along the magnetic field until almost all of their infall energy has been deposited. This is completely different from the stopping mechanism in strong fields, where $n_{max} \lesssim 1$ (e.g. Kirk & Galloway 1982, Miller, Salpeter & Wasserman 1987).

In moderate magnetic fields, $B_{12} \sim 1$, where $n_{max}$ is large, most of the incident energy of a decelerating proton is first converted into Landau excitations of the electrons it encounters along its path. An excited electron in the $n$th Landau level then decays back to the ground state primarily by successive electric dipole radiative deexcitations, thus emitting $n$ fundamental cyclotron photons via the cascade $n \to n-1$, $n-1 \to n-2$, ... $1 \to 0$. Emission of higher harmonic photons is rare mainly for two reasons: (1) the initial Landau excitations are dominated by $0 \to 1$ transitions, and (2) the excited electrons are at most only mildly relativistic — their classical gyration velocities $v_c = (2n E_B/m_e)^{1/2} \lesssim v_{ff}$ (cf. eqn. [3]) correspond to Lorentz factors $\gamma \lesssim 1.3$. (The situation is quite different for transrelativistic to very relativistic electrons [$\gamma \gtrsim 2$] in strong magnetic fields, where multipole transitions, and hence higher harmonics, are important; e.g., Harding and Preece 1987.) For the parameters used here, NSW93 found that initially only about 5% of the accretion energy is converted to cyclotron harmonics higher than the fundamental. The few higher harmonic photons produced have a large resonant optical depth ($\approx 7000(T_e/100 \text{ eV})^{-1/2} \tau_T$), and quickly spawn fundamental photons via (predominantly) dipole radiative decay of the $n > 1$ Landau levels they excite: the branching ratio of electric quadrupole to electric dipole decay rates is $\Gamma_{n \to n-2}/\Gamma_{n \to n-1} = (12/5)(E_B/m_e)(n-1) = 0.05 B_{12}(n-1)$. We do not follow higher harmonic photons in our numerical work; we assume that in a more detailed treatment, we would find that all such photons convert into fundamental photons almost immediately after birth.

We emphasize that the cyclotron photons resulting from the cascade described above are produced *nonthermally*: their initial energy, $E_B$, may be far larger than the atmospheric temperature. Ultimately, the bulk of the energy in these "primary" photons is deposited back in the atmosphere as a result of electron recoil in magnetic Compton scattering, and photon absorption; this energy emerges in the form of a low energy continuum. The cyclotron photons that manage to escape form a nonthermal high energy spectral feature.

We have used a Monte Carlo code to compute the polarized radiative transfer of cyclotron photons through the magnetized plasma (Nelson, Wang, Salpeter, &

Wasserman 1994b; hereafter NWSW). Cyclotron photons are born in the atmosphere with the profile derived in Nelson, Salpeter & Wasserman (1993). The dominant transfer effects are (1) magnetic Compton scattering which results in both angle and frequency redistribution (Comptonization), (2) polarization mode switching, and (3) absorption via inverse magnetic Bremsstrahlung. Of fundamental importance in our simulations is the low temperature of the atmosphere compared with the cyclotron energy,

$$k_B T_e = 190 \left(\frac{F_0}{10^{-4}}\right)^{1/4} M_{1.4}^{1/4} R_6^{-1/2} \text{ eV} << E_B \quad (5)$$

where $F_0 \equiv L_{accr}/L_E$. This has several consequences. First, the atmosphere may be treated as a cold plasma and we use the cold plasma polarization modes in our treatment of polarized transfer (Gnedin & Pavlov 1974, Ventura 1979, Nagel & Ventura 1983). Second, unlike the case in hot magnetic atmospheres, the Doppler core of the resonant scattering line is narrow, so that cyclotron photons can quickly escape the line core and avoid resonant absorption (Wasserman & Salpeter 1980; Wang, Wasserman & Salpeter 1988; NWSW). This greatly increases the probability that photons escape the atmosphere. Finally, despite significant energy loss to Compton recoil, the escaping cyclotron photons are much harder than the typical thermal photon, so that the emergent cyclotron emission line will be prominently displayed against the soft continuum. In the limit when mode switching is rare, the line profiles may be obtained semi-analytically (NWSW).

### 3. Emergent Line Luminosity and Spectrum

There are four parameters in our model: $M, R, B,$ and $F_0$. Without exception, we take $M_{1.4} = 1 = R_6$. We therefore construct a two-parameter family of isothermal models in $B$ and $F_0$.

The shape of the emergent photon number spectra (angle averaged and summed over polarization) is shown in Figure 1 for $B = 10^{12} G$ (panel a) and $5 \times 10^{12} G$ (panel b). For both panels, $F_0 = 10^{-4}$; the spectral shape does not depend strongly on $F_0$. The continuum is assumed to be a black body spectrum with an effective temperature determined from equation (4). To this spectrum we add the emergent cyclotron photon spectrum, with the proper weighting, computed from the Monte-Carlo transfer code. The vertical arrows in the figure show the location of the cyclotron energy, $E_B$, where line photons are born. The photon energy degradation due to electron recoil (i.e., Comptonization) is clearly evident; despite originating at energies $\sim 8 - 80$ keV (for $0.7 \lesssim B_{12} \lesssim 7$), the emergent line centers lie in the range $\sim 5 - 20$ keV.

Because of the strong energy dependence of the magnetic scattering cross sections, cyclotron photons escape in $N_{esc} \propto \tau^{2/5}$ scatters rather than the $\sim \tau^2$ scatters expected in nonmagnetic Compton scattering. Consequently, although broadened and degraded from their original energy, the accretion-induced cyclotron photons emerge as a clear line feature with $E/\triangle E \sim$ 2-4 for $0.7 \lesssim B_{12} \lesssim 7$ (NWSW). The spectrum thus consists of a soft thermal component plus a hard nonthermal cyclotron emission line superposed on the Wien tail of the underlying thermal continuum. This general form should persist even for harder (e.g., power law) continua as long as the bulk of atmospheric electrons have energies $\ll E_B$ (NWSW).

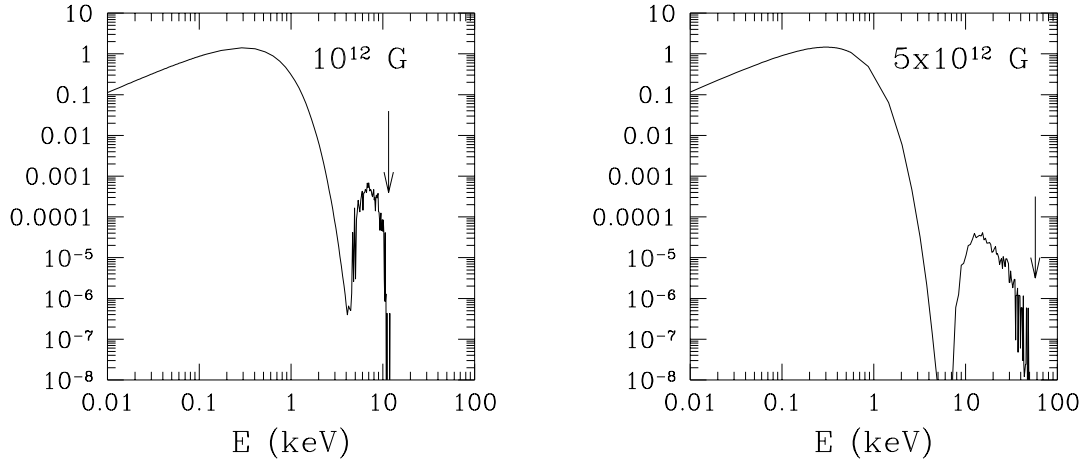

Fig. 1.— The emergent photon number spectra angle averaged, summed over polarization, and normalized to unit area from a magnetized polar cap neutron star atmosphere with (a) $B = 10^{12}\,G$, $F_0 = L_{accr}/L_E = 10^{-4}$, and $L_{line} = 0.02 L_{accr}$, and (b) $B = 5 \times 10^{12}\,G$, $F_0 = 10^{-4}$, $L_{line} = 0.01 L_{accr}$. The arrows indicate the cyclotron energy, $E_B$, where line photons are born.

Figure 2 shows the fraction of accretion luminosity that escapes in the cyclotron line, $L_{line}/L_{accr}$ as a function of magnetic field strength. The dependence on $F_0$ enters only weakly through the absorption cross section, $\sigma_{abs} \propto 1/T_e \propto F_0^{-1/4}$. At high values of $B$, few cyclotron photons are produced so the line flux falls off rapidly. The rapid fall-off at low $B$ arises from the transfer microphysics: The energy lost per scatter by photons to electron recoil along field lines is small compared to the thermal Doppler width, so that line photons become trapped inside the line core where the scattering cross section is large (cf. Wang, Wasserman & Salpeter 1988; Wasserman & Salpeter 1980) and are destroyed by absorption before they can escape. However, a significant line luminosity ($0.005 \lesssim L_{line}/L_{accr} \lesssim 0.05$) is found for a broad range of field strengths $0.7 \lesssim B_{12} \lesssim 7$. In this regime, the field is sufficiently strong to avoid line trapping, but not strong enough to quench the initial cyclotron photon production.

The soft X-rays below $\sim$ 1 keV will generally be strongly absorbed by the intervening interstellar gas but may still be visible by instruments onboard satellites such as ASCA (0.5–10 keV) and ROSAT (0.2–2.4 keV). For an assumed continuum spectral shape, the soft X-ray component can be used to obtain the line-of-sight column density and the bolometric source luminosity. However, from the soft X-ray component alone, it will be difficult to distinguish the neutron star source spectra from spectra of other types of objects such as young stellar objects and active galactic nuclei. This is especially true of the ROSAT PSPC detector (0.2–2.4 keV) where the limited energy range above $\sim 1$ keV allows for a wide range of possible continuum fits. In sharp contrast, the nonthermal hard cyclotron emission line predicted here will not be strongly absorbed and thus *serves as a unique identifying signature for strongly magnetized low luminosity accreting neutron stars*. Note that while it is possible for the ASCA detectors to observe the nonthermal component (if $B_{12} \sim 1$), according to our calculations the ROSAT detectors cannot see this feature for any field strength.

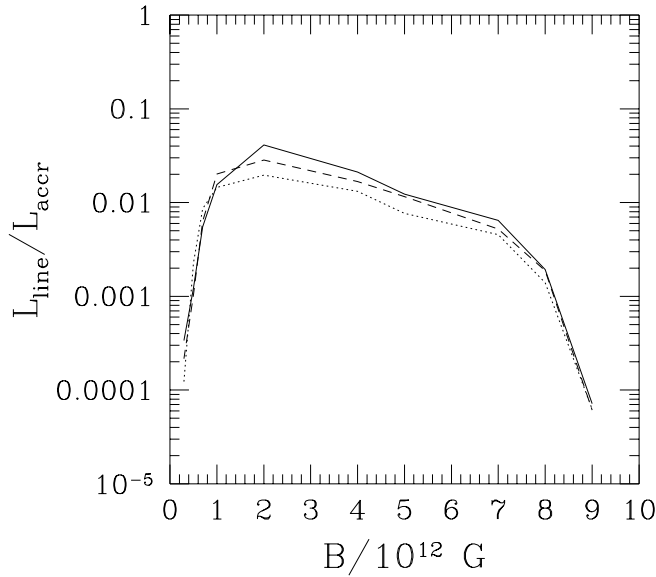

Fig. 2.— The fraction of the total accretion luminosity in the nonthermal cyclotron emission line as a function of magnetic field strength. The curves correspond to $F_0 = $ (solid)$10^{-2}$, (dashed)$10^{-4}$, and (dotted) $10^{-6}$.

## 4. Discussion

It may be possible to observe cyclotron emission line features in the spectra of highly magnetized X-ray transients during their quiescent phases, provided that some mass accretes onto these objects in spite of the propeller effect or any other factors that may inhibit low luminosity accretion. For example, consider the Be/X-ray transient pulsar 4U0115+63 which has a rotation period $P_p = 3.6$ s, is relatively nearby at $\sim 3.5$ kpc (White, Swank and Holt 1983), and has a magnetic field strength of $B_{12} \simeq 1$ as determined from cyclotron absorption features seen in its spectrum during transient outbursts (Makishima et al. 1990; Nagase et al. 1991) when $L_{transient} \sim 10^{37}$ erg s$^{-1}$. If the propeller effect were 100% efficient then accretion could only occur for $L > L_{prop} \approx 3 \times 10^{36}$ erg s$^{-1}$. Suppose that in quiescence the star would accrete from the Be star wind at a luminosity $L = fL_{prop}$ (with $f < 1$) if accretion were not inhibited at all, but, because of the propeller effect, only a fraction $\epsilon << 1$ of this amount actually accretes onto the neutron star. In this case its bolometric luminosity is $L_{accr} \sim 3\epsilon f \times 10^{36}$ erg s$^{-1}$, and, according to our calculations, the cyclotron line luminosity will be of order 50 times smaller (cf. Figure 2). A 40 kilosecond integration with the Gas Imaging Spectrometer (GIS) onboard the X-ray satellite ASCA could obtain a $4\sigma$ detection of the line feature in 4U0115+63 down to a flux level of about $4 \times 10^{-14}$ erg cm$^{-2}$ s$^{-1}$ (between 5 and 10 keV). Thus, ASCA should be able to detect a cyclotron line feature if the quiescent accretion luminosity $L_{accr} \gtrsim 3 \times 10^{33}$ erg s$^{-1}$, or $\epsilon f \gtrsim 10^{-3}$. Because of Compton recoil energy loss, this emission line will be centered at $\approx 7$ keV, well below the (fundamental) absorption-like feature seen at $\approx 12$ keV during transient outbursts. The Fe K emission line at about 6.4 keV, although nearly coincident with the cyclotron emission, should be far weaker and narrower. During outbursts, an unpulsed Fe K emission line has been seen centered at 6.4 keV with FWHM $< 0.5$ keV and $L(\text{Fe K})/L_X(7-30 \text{ keV}) \approx 9 \times 10^{-3}$ (Nagase et al. 1991), and one might be concerned that this line could be confused with the nearby cyclotron emission feature during quiescence. However, we believe that the Fe K emission line should be far weaker than the cyclotron line. This Fe K line is ascribed to the reprocessing of ionizing photons ($\gtrsim 8$ keV) by cool circumstellar wind material, leading to fluorescent emission from Fe V – X. During quiescence, the thermal production of $\gtrsim 8$ keV X-rays is negligible, so the only potential source of ionizing photons is the cyclotron feature itself. Consequently, we expect the Fe K line to be $\sim 100$ times weaker than our predicted cyclotron emission line.

The unidentified X-ray source MS0317.7-6647 is a candidate for an isolated neutron star accreting directly from the ISM (Stocke et al. 1994). The position of this source coincides with that of a diffuse cloud seen by IRAS as an infrared cirrus (Wang and Yu 1994). If the source resides within this cloud, then it is at a distance $\lesssim 100$ pc. Its ROSAT PSPC spectrum ($0.2 - 2.4$ keV) is consistent with a 170 eV black body, a line of sight column density of $N_H \simeq 4 \times 10^{21}$ cm$^{-2}$, and a total X-ray luminosity of $2 \times 10^{30}(d/100 \text{ pc})^2$ erg s$^{-1}$ (Stocke et al. 1994). The inferred accretion patch area is only $\sim 10^{-4}$ of the total surface area of a neutron star, suggesting magnetic polar cap accretion. If this source is indeed an isolated neutron star and has a surface magnetic field of $\sim 10^{12}$, then a cyclotron emission line should be present between $\sim 5$–20 keV in

– 9 –

its hard X-ray spectrum. Detection of this line would strongly suggest that this object is an isolated, magnetized neutron star.

Finally, we note that an emission-like feature at $\sim 7$ keV strikingly similar to the spectrum shown in Figure 1a has been seen in GINGA observations of the X-ray pulsar 1E 2259+586 (Koyama *et al.* 1989). The statistical significance of this feature in the pulsed spectrum is marginal, so that further observations are needed to establish its reality. If it turns out to be real, then the feature could be due to Compton degraded cyclotron emission, as predicted here. If so, the surface magnetic field strength of this pulsar is $\sim 1 \times 10^{12}$ G, and not the lower value $\sim 5 \times 10^{11}$ G previously inferred.

We thank O. Blaes, Q. D. Wang, and A. Yoshida for useful discussions, and help in deriving the parameters for line detection by ASCA. This work was supported in part by NASA grants NAGW-666, NAGW-766, NSF grants AST 91-19475, AST 91-20599, AST 93-15375, and the NSERC of Canada.

---